**Interdiffusion of Ni-Al multilayers: a continuum and molecular dynamics study**

Rong-Guang Xu[1], Michael. L. Falk[1,2,3], Timothy P. Weihs[2]

[1] Department of Physics and Astronomy, Johns Hopkins University, Baltimore MD 21218, USA

[2] Department of Materials Science and Engineering, Johns Hopkins University, Baltimore MD 21218, USA

[3] Department of Mechanical Engineering, Johns Hopkins University, Baltimore MD 21218, USA

**Abstract**

Molecular dynamics simulation of Al/Ni multilayered foils reveals a range of different reaction pathways depending on the temperature of the reaction. At the highest temperatures Fickian interdiffusion dominates the intermixing process. At intermediate temperatures Ni dissolution into the Al liquid becomes the dominant mechanism for intermixing prior to formation of the B2 intermetallic phase. At lower temperatures the B2 intermetallic forms early in the reaction process precluding both of these mechanisms. Interdiffusion and dissolution activation energies as well as diffusion prefactors are extracted from the simulations.

**1. Introduction**

Nanostructured reactive multilayer foils are composed of thousands of alternating nanometer scale layers of elements that have a large negative enthalpy of mixing. Such multilayer foils can be fabricated in the lab by magnetron sputtering or high-vacuum deposition techniques in a very controllable way. When a small but concentrated pulse of energy such as an electric spark or a thermal pulse is provided, highly exothermic, self-propagating chemical reactions can be triggered that proceed very quickly. The Ni-Al multilayer is often used as model system and as such it has been widely investigated systems both in experiment [1-5] and in continuum studies [6-16]. Recently, these multilayers have received increased attention because of their potential application as controllable, localized heat sources for joining microelectronic components without damage and as useful tools for forming near-net shape intermetallics [17-19]. A recent review can be found in Ref. [20].

In addition to being of practical importance, such materials systems provide a means to study the effect of heating rate and concentration gradients on formation reactions particularly the role of interdiffusion and nucleation. In the present work, we will focus on the interdiffusion process in Ni-Al multilayers.

Continuum modeling of the self-propagating reactions of Ni-Al reactive multilayered foils usually assumes that the diffusion process is Fickian, governed by an interdiffusion coefficient that only depends on temperature and is characterized by a single activation energy [12-16]. However, since intermetallic compounds can form during the mixing process, the reaction path could deviate significantly from this simple picture.

A number of molecular dynamics (MD) simulations have been reported in this area to date, as MD provides a promising tool to uncover the physical and chemical atomic-level mechanisms which govern the properties of such energetic materials. Earlier research includes the study of intermixing behavior for the interface of Fe/Co/Ni-Al thin film systems during deposition [21-23]. It has been used successfully to investigate the melting and crystallization in the $Al_{50}Ni_{50}$ system [24], the melting and alloying of Ni-Al nanolaminates under shock loading [25-27] and the role of pressure and misfit strain on the response of Ni-Al multilayers to thermal initiation [28-31]. Of direct relevance to this work, MD work by Baras and Politano [32] revealed the sequence of phase formation within a Ni-Al bilayer system during exothermic reactions, and work by Cherukara, Vishnu and Strachan [33] elucidated the effect of surfaces and voids on the rate of intermixing.

Here we deploy MD in the absence of free surfaces and voids to identify specific mixing mechanisms and to quantify the associated activation energies. Extracting such parameters is crucial to parameterize physically grounded predictive continuum models, which can access longer time and length scales directly comparable to experimental conditions. Here we focus on characterizing deviations from Fick's law behavior and simple Arrhenius assumptions. These deviations arise due to the nucleation of second phases and dissolution processes that can arise in lieu of diffusion.

We will first describe the details of our MD simulation and show the different phenomena corresponding to different temperature regimes. We then present a general continuum framework for analyzing the periodic bilayer Ni-Al system and quantifying the rate of interdiffusion. This is accomplished by measuring the degree of mixing $M$, which we define below, and the rate of change of $M$. We fit MD results to the analytical form and extract the diffusivity. From this data, we can determine under which conditions the assumption of Fickian interdiffusion holds and the reason for deviations when they arise. Finally we present and test a dissolution model to describe the mixing process observed in MD simulation at lower temperatures.

**II. Methodology and Data Analysis**
**Part A: Simulation Methodology**
To explore the kinetic processes in different temperature regimes, we have undertaken extensive isothermal MD simulations of interdiffusion in Ni-Al multilayers over a broad range of temperatures. In the present work, we model the Ni-Al interactions with the EAM potential developed by Y. Mishin [34], which was parameterized using experimental data combined with a large set of first-principles calculations. This potential provides an accurate description of a large variety of properties such as basic lattice properties, phonon frequencies, thermal expansion, diffusion, and equations of state for Al, Ni, and their alloys. The MD simulation is conducted using the LAMMPS software package [35].

Periodic boundary conditions are implemented in all three directions to approximate the effect of an infinite array of layers in the z direction that extend without bound in the x and y directions. The simulation domain is of size 53.0086×53.0086×80.7362Å in the x, y, and z directions, respectively. Initially, the system is comprised of compositionally pure layers, with Al atoms occupying the region 0<z<47.6344Å, and Ni atoms occupying 47.6344Å<z<80.7362Å. The initial Al/Ni bilayer used in all of the simulations presented here is composed of 8,112 Al and 8,100 Ni atoms (a nearly 1:1 stoichiometry) in a FCC lattice, arranged with [001] orientation along the z direction. The lattice constants of Ni and Al are chosen slightly off their equilibrium values so that both lattices can be accommodated within the same periodic cell. In the real experimental samples the nanostructured multilayer foils are typically comprised of alternating 10-100 nm thick layers. So the layers simulated here are at the thin layer limit of the experimental samples.

Initially the entire system is relaxed under NPT ensemble control for 5000ps at 300K and 0 GPa. The samples are then heated from 300K to desired final temperatures (from 1300K to 2000K), the temperature is held constant during the remainder of the simulation. In these simulations pressure and temperature are maintained in equilibrium with a heat bath and mechanical reservoir using a Nose-Hoover/Parinello-Rahman formalism [36-40]. Dynamic feedback between these reservoirs and the system allows the volume and the kinetic energy to fluctuate about the desired mean values as expected in an equilibrium system. All the simulations start from this sample at zero pressure. A time step of $\Delta t = 0.005$ps is used throughout this investigation. The barostatting and thermostatting time scales are chosen to be 2ps.

During the simulation, we detect the FCC-type solid solution phases and B2-type NiAl intermetallic phases that form by applying a connectivity-based structural analysis method based on the seminal work by Frank and Kasper [41]. We use a cutoff distance determined from the radial distribution function to determine whether any two atoms are neighbors. The atoms that neighbor a given atom are said to lie in its coordination surface. For any atom we are able to calculate the surface coordination number (SCN) q for each neighboring atom, defined as the number of neighbors this atom has in the coordination surface, and we define $v_q$ to be the number of neighbor atoms with SCN q. Generally speaking, the structural signature of this each atom is of the form $<v_4, v_5, v_6, v_8>$, which is sufficient to structurally characterize simple crystalline structures. It should be noted that this method fails at free surfaces and interfaces between distinct structures.

**Part B: Analysis Methodology**
Continuum models of self-propagating reactions are essentially coupled thermal and atomic transport calculations [14-16], which is based on the evolution of the section-averaged enthalpy, H,

$$\frac{\partial H}{\partial t} = \dot{Q}(M) + \nabla \cdot \vec{q}, \qquad (1)$$

where Q is the heat of reaction (or mixing) and $\vec{q}$ is the heat flux. *M* is a measure of the degree of mixing of the system, averaged over a complete bilayer. The evolution of *M* is a function of both *M* itself (because *M* depends on the composition profile) and temperature as

$$\frac{\partial M}{\partial t} = g(M,T). \tag{2}$$

To make the solution of this coupled set of equations more tractable, we assume that $g(M,T)$ is separable according to

$$\frac{\partial M}{\partial t} = f(M)D(T), \tag{3}$$

where $D(T)$ is assumed to follow an Arrhenius relationship, $D = D_0 \exp(-E/kT)$ with a single activation energy, E. In the most commonly employed version of this model, the form of *f(M)* is derived under the assumption that the net effect of interdiffusion fluxes can be modeled as a Fickian process [14-16]; that is, we assume that *M* can be described in terms of a composition profile evolving according to Fick's Law controlled by a single diffusion coefficient, *D(T)*, that is a function of temperature only. Despite this simplification, the model is observed to adequately capture macroscopic properties of the self-propagating reactions, including front velocity and temperature profile when compared to experiments in appropriate regimes [14-16].

Despite the successes of modeling that has relied on this Fickian assumption, various diffusion mechanisms can operate in nanoscale multilayers. We believe that a more sophisticated treatment of interdiffusion may be necessary, particularly when cooling rates are higher. Therefore we propose starting our analysis with the more general form of convection-diffusion equation:

$$\frac{\partial q}{\partial t} = \frac{\partial [D(q,t)\frac{\partial q}{\partial x}]}{\partial x} - v_x \frac{\partial q}{\partial x}, \tag{4}$$

combined with the continuity equation,

$$\frac{\partial \rho}{\partial t} = -\rho \frac{\partial v_x}{\partial x} - v_x \frac{\partial \rho}{\partial x}, \tag{5}$$

where *q* is the mole fraction of one of the constituent species of the multilayer. Note that mixing-induced changes in density with mole fraction will cause the material to expand or contract with time during the reaction, the size of the domain and the boundary conditions must be adjusted with time accordingly.

To simplify the problem, instead of solving Eq. (4) on the physical x axis, we will consider the diffusion along a scaled coordinate determined by the mapping $d\tilde{x} = \frac{A(t)}{N}\rho(q)dx$, where N is the total number of atoms in a single period of the multilayer, and A(t) is the cross-sectional area of the multilayer, i.e. the dimensions perpendicular to the x-axis. Note that this new coordinate is dimensionless and by construction $\Delta \tilde{x}$ across the domain is always equal to 1.

In the special case where $\frac{1}{\rho}\frac{\partial \rho}{\partial q}$ is a constant, the problem has a simple solution. Let's assume that $\rho = \rho_0 \exp(\alpha q) = \rho_B^{1-q}\rho_A^q$, i.e. q=0 corresponds to pure B. Under these circumstances we can rewrite our two equations as a single equation.

$$\frac{\partial q}{\partial t} = D(q,t)\frac{A^2(t)}{N^2}\rho^2(q)\{\frac{\partial^2 q}{\partial \tilde{x}^2}+(2\alpha+\frac{1}{D}\frac{\partial D}{\partial q})(\frac{\partial q}{\partial \tilde{x}})^2\} \quad (6)$$

At this point we can observe that under a particular assumption, namely that $D(q,t) = D_B(t)\exp(-2\alpha q) = D_B(t)(\frac{\rho_B}{\rho_A})^{2q}$, Eq. (6) simplifies significantly to a form analogous to normal diffusion equation,

$$\frac{\partial q}{\partial t} = \hat{D}(t)\frac{\partial^2 q}{\partial \tilde{x}^2} \quad (7)$$

where $\hat{D}(t) = \frac{A^2(t)}{N^2}\rho_A^2 D_A(t) = \frac{A^2(t)}{N^2}\rho_B^2 D_B(t)$. While this solution is not fully general, it captures essential aspects of the change in lattice parameter while remaining analytically tractable, and thereby presents advantages for analyzing our simulation results.

Since the solution must be periodic we can reduce our expression for the concentration to a Fourier series. In doing so we will immediately de-dimensionalize lengths in terms of the width of the bilayer using mapping equation. Here we will assume that we are mixing two pure metals and we will use $\xi$ to denote the mole fraction of species A in a full periodic layer of the structure. We will also de-dimensionalize the concentration by the initial concentration in the species of interest,

$$q(\tilde{x},t) = \xi + \sum_{n=1}^{\infty} a_n(t)\sin(2\pi n\tilde{x}) + \sum_{n=1}^{\infty} b_n(t)\cos(2\pi n\tilde{x}) \quad (8)$$

To satisfy Eq. (7) it must be true that

$$\frac{\partial a_n}{\partial t} = -4\pi^2 n^2 \hat{D}(t)a_n; \quad \frac{\partial b_n}{\partial t} = -4\pi^2 n^2 \tilde{D}(t)b_n \quad (9)$$

which leads to the solutions

$$a_n = a_n^0 \exp(-4\pi^2 n^2 \int \hat{D} dt); \quad b_n = b_n^0 \exp(-4\pi^2 n^2 \int \hat{D} dt) \quad (10)$$

Eq. (10) justifies the introduction of the dimensionless parameter that characterizes the extent of the reaction which we will call $\tau$ defined such that

$$\tau = \int \hat{D} dt \quad (11)$$

We can calculate the initial values of the a and b coefficients directly as

$$a_n^0 = 2\int_{-\xi/2}^{\xi/2} \sin(2\pi nx)\,dx = 0$$

$$b_n^0 = 2\int_{-\xi/2}^{\xi/2} \cos(2\pi nx)\,dx = \frac{2}{\pi n}\sin(\pi n\xi) \quad . \tag{12}$$

This results in the following expressions for the time dependent coefficients

$$a_n = 0; \quad b_n = \frac{2}{\pi n}\sin(\pi n\xi)\exp(-4\pi^2 n^2 \tau) \quad . \tag{13}$$

In order to reduce the order of the model we define a scalar measure that quantifies the degree of mixing,

$$M = \frac{1 - \frac{1}{\xi}\int_{-\hat{w}/2}^{\hat{w}/2} q\frac{A(t)}{N}\rho(q(x))\,dx}{1-\xi} = \frac{1 - \frac{1}{\xi}\int_{-\xi/2}^{\xi/2} q\,d\tilde{x}}{1-\xi} \quad, \tag{14}$$

where $\hat{w}$ is defined by the condition $\xi = \int_{-\hat{w}/2}^{\hat{w}/2} \frac{A(t)}{N}\rho(q(x))\,dx$ or, equivalently since the solution is symmetric $\frac{d\xi}{d\hat{w}} = \frac{A(t)}{N}\rho(q(\hat{w}))$. Note that instead of integrating the number fraction over a fixed region of space, we are counting the initial number of atoms in the unmixed A layer. We then measure mixing by evaluating the number fraction within the equivalent number of atoms centered around the middle of the A layer although the spatial extent of that layer may be expanding or contracting during interdiffusion.

A formal computation of $M$ can be obtained by integrating Eq. (8).

$$M = 1 - \frac{2}{\pi^2 \xi(1-\xi)}\sum_{n=1}^{\infty}\frac{1}{n^2}\sin^2(\pi n\xi)\exp(-4\pi^2 n^2 \tau) \quad . \tag{15}$$

We can also calculate the rate of change of M at any time via the equation

$$\dot{M} = \frac{8\hat{D}(t)}{\xi(1-\xi)}\sum_{n=1}^{\infty}\sin^2(\pi n\xi)\exp(-4\pi^2 n^2 \tau) \quad . \tag{16}$$

We can see that for this case we have two asymptotic limits. In the limit of small $\tau$ we can convert the sums into integrals and obtain

$$M \approx \frac{2}{\pi^2 \xi(1-\xi)}\sum_{n=1}^{\infty}\frac{1}{n^2}\sin^2(\pi n\xi)[1-\exp(-4\pi^2 n^2 \tau)]$$

$$\approx \frac{4\sqrt{\tau}}{\pi\xi(1-\xi)}\int_0^{\infty}\frac{\sin^2(\frac{\xi u}{2\sqrt{\tau}})}{u^2}(1-e^{-u^2})\,du = \frac{2}{\xi(1-\xi)}\sqrt{\frac{\tau}{\pi}} \quad, \tag{17}$$

$$\dot{M} \approx \frac{4\hat{D}}{\pi\xi(1-\xi)\sqrt{\tau}}\int_0^{\infty}\sin^2(\frac{\xi u}{2\sqrt{\tau}})e^{-u^2}\,du = \frac{\hat{D}}{\xi(1-\xi)\sqrt{\pi\tau}} \quad . \tag{18}$$

This results in a relationship of the form

$$\dot{M} = \frac{2}{\pi\xi^2(1-\xi)^2}\hat{D}M^{-1} \quad . \tag{19}$$

In the large $\tau$ limit the first term in the series dominates and

$$M = 1 - \frac{2}{\pi^2 \xi(1-\xi)} \sin^2(\pi\xi) \exp(-4\pi^2\tau), \tag{20}$$

$$\dot{M} = \frac{8\hat{D}(t)}{\xi(1-\xi)} \sin^2(\pi\xi) \exp(-4\pi^2\tau). \tag{21}$$

So we can write that in this limit

$$\dot{M} = 4\pi^2 \hat{D}(1-M). \tag{22}$$

Here $\hat{D}(t) = \frac{A^2(t)}{N^2} \rho_A \rho_B D_{eff}(t)$, where we define $D_{eff}(t)$ as the "effective diffusivity" since this is the diffusivity one would expect assuming a Fickian process. Of course, in actuality, the process may deviate significantly from this Fickian ideal, but expressing intermixing rates in terms of this "effective diffusivity" will be useful for comparing these various scenarios. Furthermore this quantity can be extracted directly from the graph of $\dot{M}$ vs $M$.

### III. Results and Data Analysis
There are three different scenarios we observe in our MD simulations corresponding to different behavior during mixing: high temperature (1700K to 2000K), intermediate temperature (1400K to 1600K) and low temperature (1300K and 1350K) regimes.

At high temperatures, the Al layer melts and Ni atoms dissolve into the liquid Al layer, forming a homogenous alloy liquid at the final stage. The dissolution process is very quick and no solid solution or B2 NiAl phases form during the reaction. Regarding the mixing, the rate of change of *M* as a function of *M* in MD simulation is nearly linear as shown in Figure 1. From this we can extract the interdiffusivity from the slope using the large time asymptotic solution, Eq. (19), as shown in Figure 2(a). Note that Figure 2(b) also includes data from temperatures 1500K and 1600K in the late stages when these systems also exhibit constant effective interdiffusion coefficients as shown in Figure 3(a). We can see that interdiffusivity for high temperatures follows a simple Arrhenius relationship very closely, justifying the assumption of an interdiffusion coefficient that is nearly independent of composition and which arises from a single kinetic process. From the data, the activation energy is 35.77kJ/mole (0.3709 eV) and $D_0$=17.617 Å/ps.

At intermediate temperatures, a solid solution forms at the interface during the dissolution process before the whole system melts. The amount of solid solution formed at the interface increases with time as shown in Figure 3(b). As mixing proceeds the effective interdiffusion coefficient decreases until the solid solution melts. At that time, the effective interdiffusion coefficient increases and approaches the value predicted by the kinetic parameters measured at high temperature as detailed in the previous paragraph. This qualitatively different mixing sequence appears to arise due to the existence of the boundary layer formed by the solid solution. The mixing behavior becomes much more complicated and can no longer be described in the framework of a single simple Fickian diffusion process.

At lower temperatures, 1300K and 1350K illustrated in Figure 4, the solid solution is again observed to form at the interface, and, consequently, the effective interdiffusion coefficient decreases dramatically. Then the solid solution dissolves, the effective interdiffusion coefficient increases, but almost immediately thereafter we observe the nucleation of a B2 phase as shown in Figure 5. The effective interdiffusion coefficient decreases as the B2 phase quickly grows to encompass nearly the entire system.

Below 1600K complete mixing is never achieved due to the formation of intermetallic phases. We observe that the actual mixing process at these temperatures involves Ni atoms dissolving into the molten Al layer. As the Ni concentration increases, the rate of dissolution decreases, simultaneously Al atoms diffuse into the solid Ni layer. Thus there are two processes occurring during the mixing. A model for this behavior has been discussed previously in the literature in the context of reactive bilayers [10-11] and solders [42-43].

During pure dissolution, according to Noyes and Whitney's law [44], the rate of dissolution is proportional to the difference between the instantaneous concentration, $q$ at time t, and the saturation solubility $q_{Ni}^{S}$. Thus the dissolution process should follow 1$^{st}$ order kinetics according to the expression

$$\frac{dq_{Ni}^{dis}(t)}{dt} = K(q_{Ni}^{S} - q_{Ni}^{dis}(t)) \qquad (23)$$

The solution of this equation is

$$q_{Ni}^{dis}(t) = q_{Ni}^{S}(1 - e^{-Kt}) \qquad (24)$$

Here $q_{Ni}^{dis}(t)$ is mole fraction of Ni atoms in molten Al layer due to pure dissolution.

$$q_{Ni}^{dis}(t) = \frac{\rho_{Ni} l(t)}{\rho_{Ni} l(t) + \rho_{Al} w_0} = \frac{l(t)}{\alpha + l(t)}, \qquad (25)$$

where $\alpha = \frac{\rho_{Al}}{\rho_{Ni}} w_0$ and $w_0$ is the half length of the molten Al layer at time t=0.

Then

$$l(t) = \frac{\rho_{Al}}{\rho_{Ni}} \frac{1 - e^{-Kt}}{\frac{1 - q_{Ni}^{S}}{q_{Ni}^{S}} + e^{-Kt}} w_0 = \alpha \frac{1 - e^{-Kt}}{\frac{1 - q_{Ni}^{S}}{q_{Ni}^{S}} + e^{-Kt}}. \qquad (26)$$

At the same time, Al atoms also diffuse into Ni layer as described by the diffusion equation,

$$\frac{\partial q_{Al}}{\partial t} = D_{Al} \frac{\partial^2 q_{Al}}{\partial x^2},\tag{27}$$

with the initial and boundary conditions

$$q_{Al}(x, t=0) = 0;\ q_{Al}(x = l(t), t) = q_{Al}^S.\tag{28}$$

Here $q_{Al}^S$ is the mole fraction of Al in the Ni solid layer and $q_{Al}^S = 1 - q_{Ni}^S$.

This is known as the Stefan problem, but a case in which the boundary condition in Eq. (28) is defined on the non-stationary, moving interface $x = l(t)$ between the melt Al and metal Ni. This problem only has known analytical solutions for cases in which $l(t) = \alpha\sqrt{t}$ [45-46]. However, it is clear that the initial flux due to Al diffusion into the solid Ni layer is much smaller than the flux due to Ni dissolution into the molten Al. So, we can neglect the effect of the Al diffusion on the velocity of the moving interface.

From this analysis we can compute the mixing measure,

$$M(t) = \frac{q_{Ni}(t)}{1-\zeta} = 2 q_{Ni}(t) \cong 2 q_{Ni}^{dis}(t) = 2 q_{Ni}^S (1 - e^{-Kt}),\tag{29}$$

$$\dot{M}(t) = \frac{\dot{q}_{Ni}(t)}{1-\zeta} = 2 \dot{q}_{Ni}(t) \cong 2 \dot{q}_{Ni}^{dis}(t) = 2 q_{Ni}^S K e^{-Kt} = K(2 q_{Ni}^S - M) = -K(M - 2 q_{Ni}^S),\tag{30}$$

Eq. (30) can adequately describe the $\dot{M}(t)$ vs $M$ curve from our MD data as shown in Figures 6 and 7. We can extract K and $q_{Ni}^S$, respectively for different temperatures, and, as shown in Figure 8, we find a reasonably good Arrhenius fit for the dissolution rate constant over the applicable temperature region such that

$$K(T) = K_0 \exp(-\frac{E}{RT}).\tag{31}$$

We obtain an activation barrier of 101 kJ/mol (1.046 eV), indicating that the same dissolution mechanism dominates the mixing process over the low temperature range from 1300K to 1600K.

**IV. Summary and conclusions**

We have carried out extensive isothermal MD simulation of interdiffusion in Ni-Al multilayers over a broad range of temperatures to explore the kinetic process in different temperature regimes. We fit our MD results to a Fickian interdiffusion model and extract the effective interdiffusion coefficient. We conclude that at high temperature the assumption of Fickian interdiffusion holds, but that this assumption breaks down at lower temperature due to the formation of an intermetallic phase at the boundary. Finally we construct a dissolution model to describe the mixing process observed in MD simulations at lower temperatures. This model is found to satisfactorily describe the process observed at lower temperatures. Activation barriers for both processes have been extracted from the data and these indicate that the activation barrier is three times as high at lower temperature due to the intermetallic boundary layer.

## V. Acknowledgement

This research was supported by Department of Energy through Grant No. DE-SC002509. We thank Todd Hufnagel and Omar Knio for helpful discussions and for sharing information regarding their experimental and computational investigations. Simulations were performed at the Homewood High Performance Cluster with facilities supported under Grant No. OCI 0963185.

**References**
[1] E. Ma, C. V. Thompson, L. A. Clevenger, and K. N. Tu, Applied Physics Letters **57**, 1262 (1990).
[2] E. Ma, C. V. Thompson, and L. A. Clevenger, Journal Of Applied Physics **69**, 2211 (1991).
[3] A. S. Edelstein, R. K. Everett, G. Y. Richardson, S. B. Qadri, E. I. Altman, J. C. Foley, and J. H. Perepezko, Journal Of Applied Physics **76**, 7850 (1994).
[4] K. Barmak, C. Michaelsen, and G. Lucadamo, Journal Of Materials Research **12**, 133-146 (1997).
[5] K. J. Blobaum, D. Van Heerden, A. J. Gavens, and T. P. Weuhs, Acta Materialia **51**, 3871-3884 (2003).
[6] A. B. Mann, A. J. Gavens, M. E. Reiss, D. Van Heerden, G. Bao, and T. P. Weihs, Journal Of Applied Physics **82**, 1178 (1997).
[7] S. Jayaraman, O. M. Knio, A. B. Mann, and T. P. Weihs, Journal Of Applied Physics **86**, 800 (1999).
[8] S Jayaraman, A. B. Mann, M. Reiss, T. P. Weihs, and O. M. Knio, Combustion And Flame **124**, 178-194 (2001).
[9] E. Besnoin, S. Cerutti, O. M. Knio, and T. P. Weihs, Journal Of Applied Physics **92**, 5474 (2002).
[10] A. Makino, Combustion and Flame 134, 273-288 (2003).
[11] A. Makino, Proceedings of the Combustion Institute 31, 1813-1820 (2007).
[12] B. B. Khina, Journal Of Applied Physics **101**, 063510 (2007).
[13] R. Knepper, M. R. Snyder, G. Fritz, K. Fisher, O. M. Knio, and T. P. Weihs, Journal Of Applied Physics **105**, 083504 (2009).
[14] M. Salloum and O. M. Knio, Combustion And Flame **157**, 288-295 (2010).
[15] M. Salloum and O. M. Knio, Combustion And Flame **157**, 436-445 (2010).
[16] M. Salloum and O. M. Knio, Combustion And Flame **157**, 1154-1166 (2010).
[17] J. Wang, E. Besnoin, A. Duckham, S. J. Spey, M. E. Reiss, O. M. Knio, M. Powers, M. Whitener, and T. P. Weihs, Applied Physics Letters **83**, 3987 (2003).
[18] J. Wang, E. Besnoin, A. Duckham, S. J. Spey, M. E. Reiss, O. M. Knio, and T. P. Weihs, Journal Of Applied Physics **95**, 248 (2004).

[19] J. Wang, E. Besnoin, O. M. Knio, and T. P. Weihs, Journal Of Applied Physics **97**, 114307 (2005).
[20] A. S. Rogachev, Russ. Chem. Rev. **77**, 21 (2008).
[21] S-P Kim, Y-C Chung, S-C Lee, K-R Lee and K-H Lee, Journal Of Applied Physics **93**, 8564 (2003).
[22] S-G Lee, S-P Kim, K-R Lee and Y-C Chung, Journal Of Magnetism and Magnetic Materials **286**, 394 (2005).
[23] C-Y Chung and Y-C Chung, Materials Letter, **60**, 1063 (2006).
[24] B. J. Henz, T. Hawa and M. Zachariah, Europhysics Letter, **81**, 58001 (2008).
[25] S. Zhao, T. C. Germann and A. Strachan, J. Chem. Phys. **125**, 164707 (2006).
[26] S. Zhao, T. C. Germann and A. Strachan, Phys. Rev. B. **76**, 014103 (2007).
[27] S. Zhao, T. C. Germann and A. Strachan, Phys. Rev. B. **76**, 014105 (2007).
[28] N. S. Weingarten, W. D. Mattson, and B. M. Rice, Journal Of Applied Physics **106**, 063524 (2009).
[29] N. S. Weingarten, W. D. Mattson, A. D. Yau, T. P. Weihs, and B. M. Rice, J. App. Phys. **107**, 093517 (2010).
[30] J. C. Crone, J. Knap, P. W. Chung and B. M. Rice, Appl. Phys. Lett. **98**, 141910 (2011).
[31] N. S. Weingarten, and B. M. Rice, J. Phys.: Condesn. Matter **23** 275701 (2011).
[32] F. Baras and O. Politano, Phys. Rev. B. **84**, 024113 (2011).
[33] M. J. Cherukara, K. G. Vishnu, and A. Strachan, Phys. Rev. B. **86,** 075470 (2012).
[34] Y Mishin, Acta Materialia **52**, 1451 (2004).
[35] See [http://lammps.sandia.gov/];S. Plimpton, J. Comput. Phys. **117**, 1 (1995).
[36] S. Nose, Molecular Physics **52**, 255 (1984).
[37] S. Nose, J. Chem. Phys. **81**, 511 (1984).
[38] W. G. Hoover, Phys. Rev. A. **31**,1695 (1985).
[39] M. Parrinello, and A. Rahman, Phys. Rev. Lett. **45**, 1196 (1980).
[40] M. Parrinello, and A. Rahman, Journal Of Applied Physics **52**, 7182 (1981).
[41] F. C. Frank, and J. S. Kasper, Acta Crystallographica **11**, 184 (1958).
[42] J. Drapala, P. Kubicek, J. Vrestal and M. Losertova, Defect and Diffusion Forum **263**, 231 (2007).
[43] J. Drapala, P. Kubicek, P. Harcuba, V. Vodarek, P. Jopek, D. Petlak, G. Kostiukova and K. Konecna, Defect and Diffusion Forum **322**, 41 (2012).
[44] A. A. Noyes and W. R. Whitney, J. Am. Chem. Soc. **19**, 930 (1897)
[45] P. Kubicek and L. Mrazek, Czech. J. Phys. **46**, 509 (1996)
[46] P. Kubicek and L. Mrazek, Czech. J. Phys. **46**, 931 (1996)

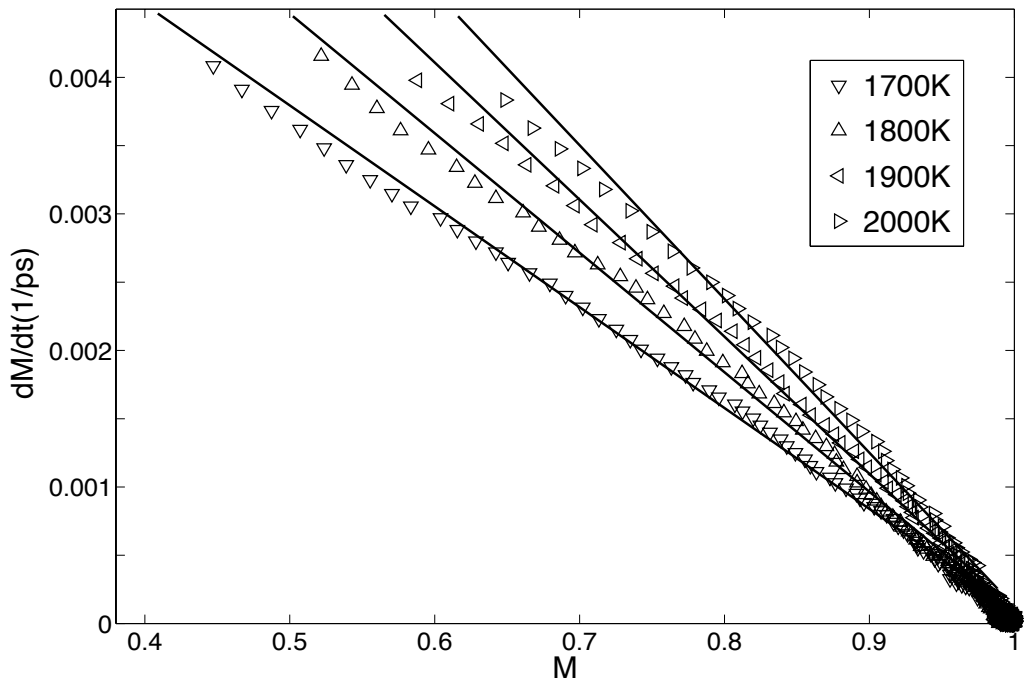

Figure 1: At high temperature, the rate of change of the degree of mixing as a function of the degree of mixing in MD simulation follows a straight line as expected for Fickian interdiffusion. The slope of these curves is used to extract the effective interdiffusion coefficient.

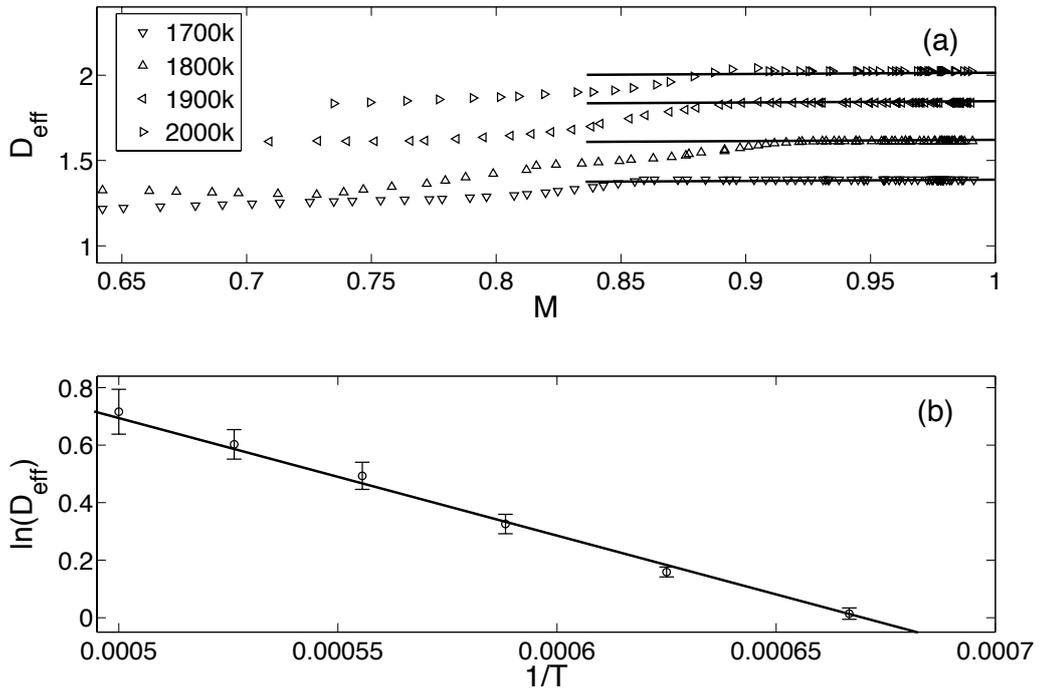

Figure 2: The top graph (a) shows that at high temperature, the effective interdiffusion coefficient is almost constant, independent of the degree of mixing. The straight lines are fits to represent the effective interdiffusion coefficient in the nearly mixed limit. The bottom graph (b) shows that an Arrhenius law governs interdiffusion. From this data we can extract the effective activation energy of 35.77 kJ/mole (0.3709eV) and the prefactor that governs interdiffusion, $D_0$=17.617 Å/ps.

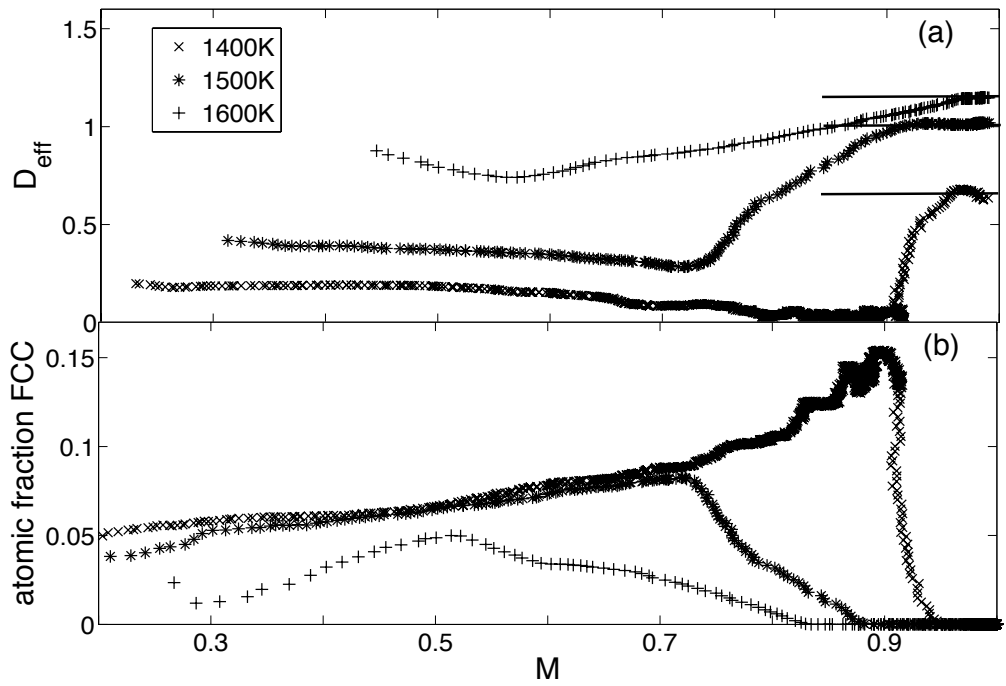

Figure 3: At intermediate temperatures, the amount of solid solution at the interface increases and the rate of interdiffusion decreases. When the solid solution melts again, the effective interdiffusion coefficient increases and approaches its asymptotic value represented by the straight line fit.

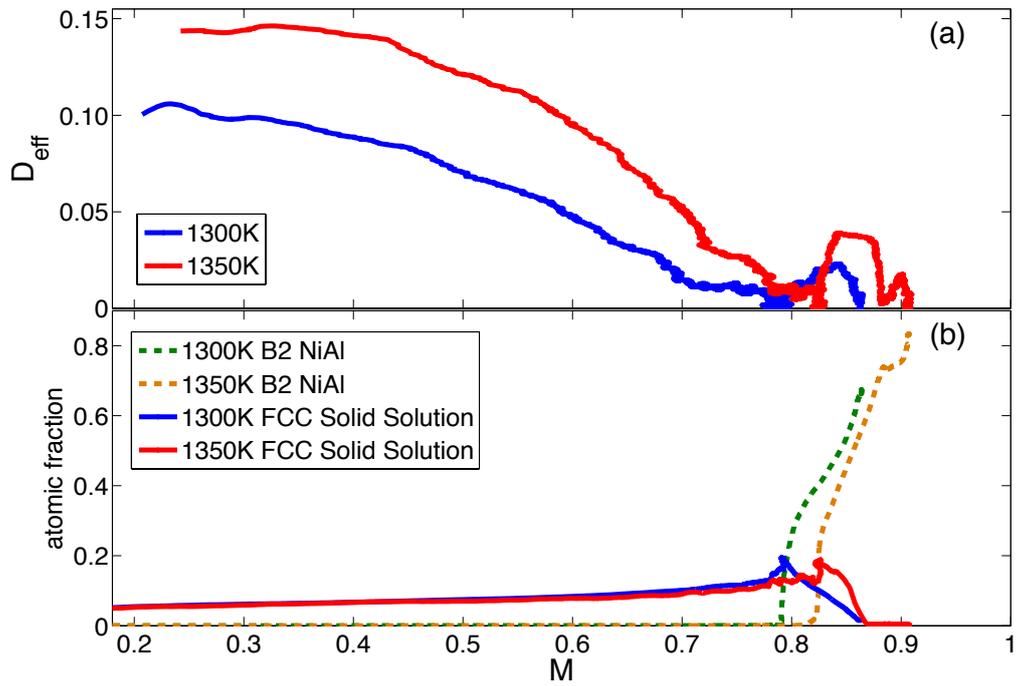

Figure 4: At 1300K and 1350K, when the amount of solid solution formed at the interface increases, the rate of interdiffusion decreases. When the solid solution starts to dissolve, the rate of interdiffusion increases correspondingly. Finally the B2 phase nucleates and rate of interdiffusion decreases again as B2 phase grows.

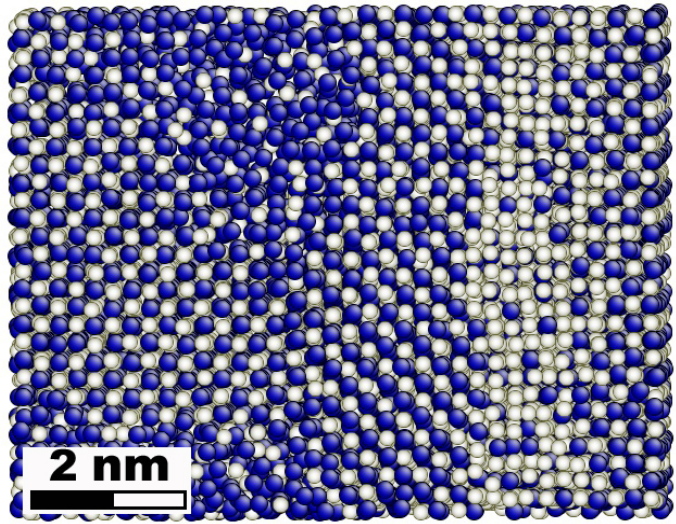

Figure 5: For simulations performed at 1300K the B2 NiAl phase nucleates and grows to encompass nearly the entire system. Al and Ni are shown as blue and white spheres, respectively.

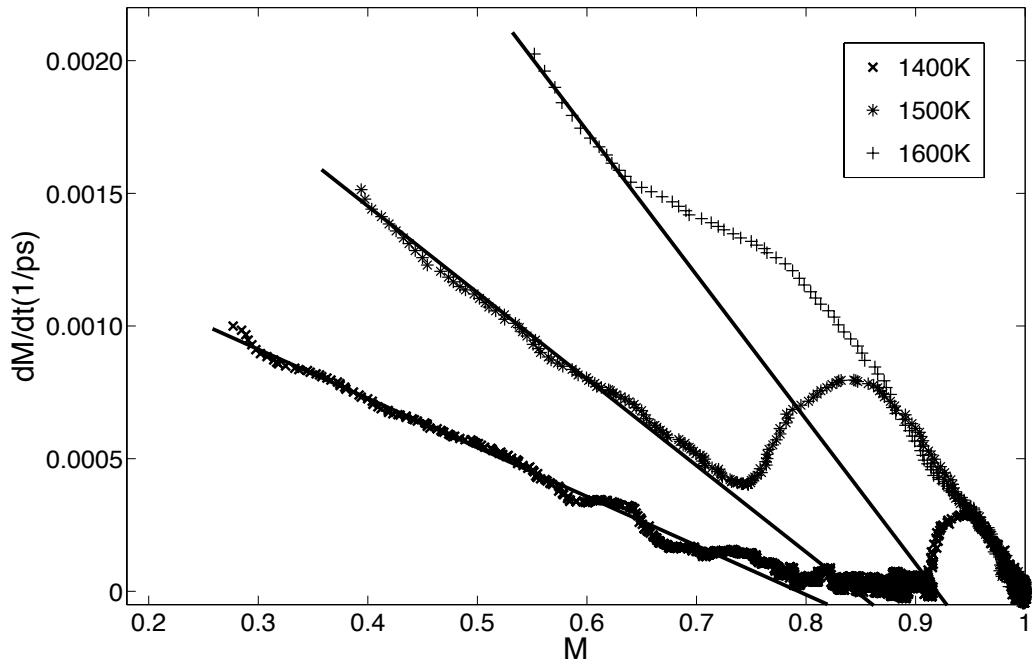

Figure 6: At intermediate temperature, the rate of change of the degree of mixing as a function of the degree of mixing in MD simulation initally decreases linearly corresponding to dissolution process. This is followed by a kink when Ni layer totally melts and finally approaches its asymptotic value.

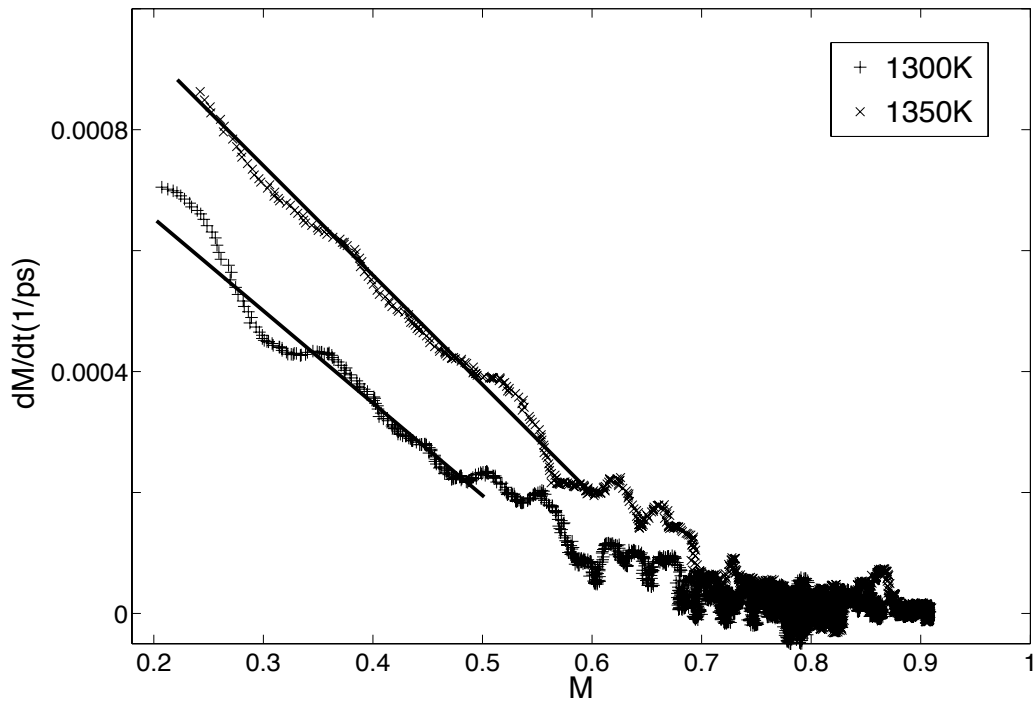

Figure 7: At low temperature, the rate of change of the degree of mixing as a function of the degree of mixing in the simulations initially decreases linearly corresponding to dissolution process, followed by formation of solid solution and nucleation of B2 NiAl phase.

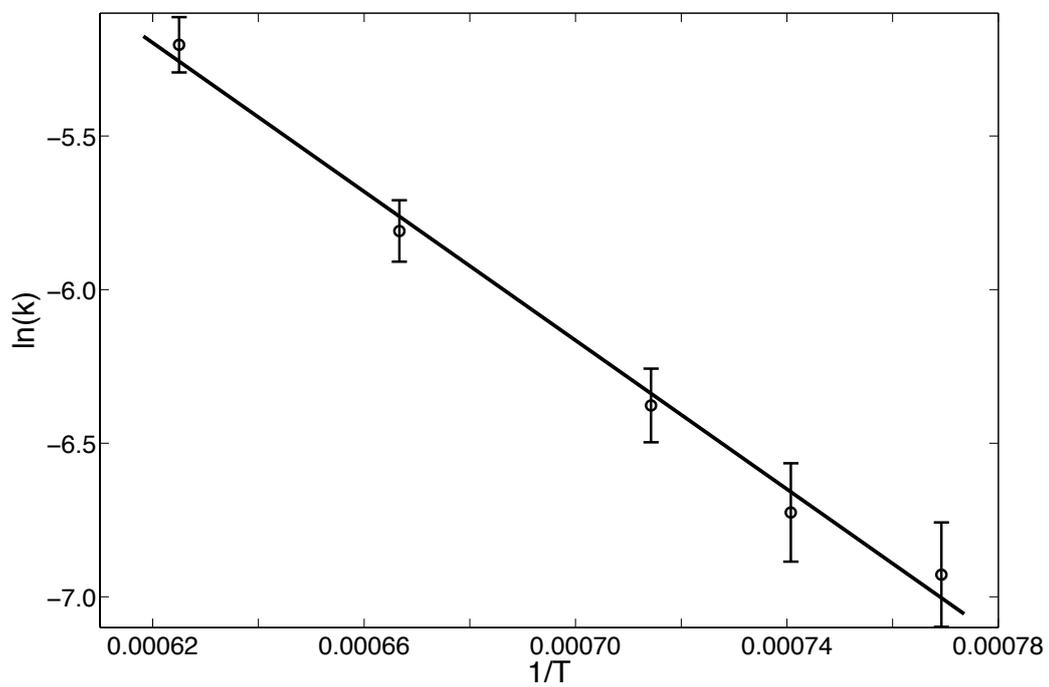

Figure 8: Arrhenius fit for dissolution rate constant reveals an activation barrier of 101 kJ/mol (1.046 eV).